\documentclass[aps,prl,superscriptaddress,twocolumn,showpacs,amsmath,amssymb,floatfix]{revtex4}

\usepackage{graphicx}
\usepackage{dcolumn}
\usepackage{bm}

\newcommand{\UAz}{University of Arizona, Tucson, Arizona 85721}
\newcommand{\UCLA}{University of California at Los Angeles, Los Angeles,
                    California 90095}
\newcommand{\Campinas}{Universidade Estadual de Campinas, Campinas,
                       Brazil 13083-970}
\newcommand{\EFI}{The Enrico Fermi Institute, The University of Chicago,
                  Chicago, Illinois 60637}
\newcommand{\UB}{University of Colorado, Boulder, Colorado 80309}
\newcommand{\ELM}{Elmhurst College, Elmhurst, Illinois 60126}
\newcommand{\FNAL}{Fermi National Accelerator Laboratory,
                   Batavia, Illinois 60510}
\newcommand{\Osaka}{Osaka University, Toyonaka, Osaka 560-0043 Japan}
\newcommand{\Rice}{Rice University, Houston, Texas 77005}
\newcommand{\SaoPaulo}{Universidade de S\~ao Paulo, S\~ao Paulo, Brazil 05315
-970}
\newcommand{\UVa}{The Department of Physics and Institute of Nuclear and
                  Particle Physics, University of Virginia,
                  Charlottesville, Virginia 22901}
\newcommand{\UW}{University of Wisconsin, Madison, Wisconsin 53706}

\begin{document}


\title{Search for Lepton Flavor Violating Decays of the Neutral Kaon}

\affiliation{\UAz}
\affiliation{\UCLA}
\affiliation{\Campinas}
\affiliation{\EFI}
\affiliation{\UB}
\affiliation{\ELM}
\affiliation{\FNAL}
\affiliation{\Osaka}
\affiliation{\Rice}
\affiliation{\SaoPaulo}
\affiliation{\UVa}
\affiliation{\UW}

\author{E.~Abouzaid}      \affiliation{\EFI}
\author{M.~Arenton}       \affiliation{\UVa}
\author{A.R.~Barker}      \altaffiliation[Deceased.]{ } \affiliation{\UB}
\author{L.~Bellantoni}    \affiliation{\FNAL}
\author{A.~Bellavance}    \affiliation{\Rice} 
\author{E.~Blucher}       \affiliation{\EFI}
\author{G.J.~Bock}        \affiliation{\FNAL}
\author{E.~Cheu}          \affiliation{\UAz}
\author{R.~Coleman}       \affiliation{\FNAL}
\author{M.D.~Corcoran}    \altaffiliation[To whom correspondence should be addressed] {  }
\affiliation{\Rice}
\author{B.~Cox}           \affiliation{\UVa}
\author{A.R.~Erwin}       \affiliation{\UW}
\author{C.O.~Escobar}     \affiliation{\Campinas}  
\author{A.~Glazov}        \affiliation{\EFI}
\author{A.~Golossanov}    \affiliation{\UVa} 
\author{R.A.~Gomes}       \affiliation{\Campinas}
\author{P. Gouffon}       \affiliation{\SaoPaulo}
\author{Y.B.~Hsiung}      \affiliation{\FNAL}
\author{D.A.~Jensen}      \affiliation{\FNAL}
\author{R.~Kessler}       \affiliation{\EFI}
\author{K.~Kotera}        \affiliation{\Osaka}
\author{A.~Ledovskoy}     \affiliation{\UVa}
\author{P.L.~McBride}     \affiliation{\FNAL}

\author{E.~Monnier}
   \altaffiliation[Permanent address ]{C.P.P. Marseille/C.N.R.S., France}
   \affiliation{\EFI}  

\author{H.~Nguyen}       \affiliation{\FNAL}
\author{R.~Niclasen}     \affiliation{\UB}
\author{D.G.~Phillips~II} \affiliation{\UVa}
\author{H.~Ping}         \affiliation{\UW}  
\author{E.J.~Ramberg}    \affiliation{\FNAL}
\author{R.E.~Ray}        \affiliation{\FNAL}

\author{M.~Ronquest}     \affiliation{\UVa}
\author{E.~Santos}       \affiliation{\SaoPaulo}
\author{W.~Slater}       \affiliation{\UCLA}
\author{D.~Smith}        \affiliation{\UVa}
\author{N.~Solomey}      \affiliation{\EFI}
\author{E.C.~Swallow}    \affiliation{\EFI}\affiliation{\ELM}
\author{P.A.~Toale}      \affiliation{\UB}
\author{R.~Tschirhart}   \affiliation{\FNAL}
\author{Y.W.~Wah}        \affiliation{\EFI}
\author{J.~Wang}         \affiliation{\UAz}
\author{H.B.~White}      \affiliation{\FNAL}
\author{J.~Whitmore}     \affiliation{\FNAL}
\author{M.~J.~Wilking}      \affiliation{\UB}
\author{B.~Winstein}     \affiliation{\EFI}
\author{R.~Winston}      \affiliation{\EFI}
\author{E.T.~Worcester}  \affiliation{\EFI}
\author{M.~Worcester}    \affiliation{\EFI}
\author{T.~Yamanaka}     \affiliation{\Osaka}
\author{E.~D.~Zimmerman} \affiliation{\UB}
\author{R.F.~Zukanovich} \affiliation{\SaoPaulo}

\begin{abstract}

   The Fermilab KTeV experiment has searched for lepton-flavor-violating decays of the 
$K_L$ meson in three decay modes. 
We observe no events in the signal region for any of the modes studied, and we set 
the following upper limits for their branching ratios at the 90\% CL: $BR(K_L 
\rightarrow \pi^0 \mu^{\pm} e^{\mp}) < 7.56 \times 10^{-11}$; $BR(K_L \rightarrow 
\pi^0 \pi^0 \mu^{\pm} e^{\mp}) < 1.64 \times 10^{-10}$;
$BR(\pi^0 \rightarrow  \mu^{\pm} e^{\mp}) < 3.59 \times 10^{-10}$. This result represents a 
factor of 82 improvement in the branching ratio limit for $K_L \rightarrow \pi^0 \mu e$ and  
is the first reported limit for $K_L \rightarrow \pi^0 \pi^0 \mu^{\pm} e^{\mp}$.

\end{abstract}

\maketitle

In the Standard Model of particle physics lepton-flavor-violating (LFV) 
decays are possible with non-zero neutrino masses and mixing, but the rates for such 
decays are far beyond the reach of any current experiment \cite{landsberg}. Therefore, 
the observation 
of LFV decays would be an indication of new physics.  Many scenarios for physics 
beyond the Standard Model allow LFV decays.
 Supersymmetry \cite{ellis}, new massive gauge bosons \cite 
{landsberg,cahn}, and 
Technicolor \cite {technicolor} all can lead to LFV decays 
which might be within reach of current experiments. Searches in $K_L$ decays are  
complementary to searches in the charged lepton sector,  since $K_L$ decays 
probe the $s \rightarrow d \mu e$ transition \cite{landsberg}.

In this letter we report on searches for three LFV processes in the KTeV experiment at 
Fermilab. We present improved 
limits on the decays $K_L \rightarrow \pi^0 \mu^{\pm} e^{\mp}$ and $\pi^0 
\rightarrow 
\mu^{\pm} e^{\mp}$  (tagged from $K_L \rightarrow \pi^0 \pi^0 \pi^0$), and we report 
the first limit on the decay $K_L \rightarrow \pi^0 \pi^0 \mu^{\pm} 
e^{\mp}$.

  The KTeV E799-II experiment at Fermilab took data in 1997 and 1999.  The 
combined 
results from both periods are presented here. 
 The KTeV beam was produced by 800 GeV/c protons from the Tevatron which were 
directed onto a BeO target and collimators 
to create two nearly-parallel $K_L$ beams. The beams entered a 65m long vacuum tank
which defined the fiducial volume for accepted decays.
 
Charged particles were detected by two pairs of drift chambers
separated by an analysis magnet that provided a transverse momentum kick 
of  either 0.250 GeV/c (for the 1997 data) or 0.150 GeV/c (for the 1999 data).
 Discrimination between charged pions and electrons was 
provided by a set of transition radiation detectors (TRDs) behind the 
last drift chamber. 
Downstream of the TRDs were two planes of trigger hodoscopes, 
 followed by a CsI electromagnetic 
calorimeter, which had an energy resolution 
$\sigma(E)/E = 0.45\% \oplus 2\% /\sqrt{E(GeV)}$.
 The calorimeter provided powerful electron/pion discrimination based on the
ratio of energy as measured in the calorimeter ($E$) to momentum as 
measured
in the spectrometer ($p$), or $E/p$. The lateral shower shape in the 
calorimeter
 provided additional electron/pion discrimination.
The CsI calorimeter had two beam holes to allow the undecayed beam particles to 
pass through. A Beam Anti (BA) calorimeter covered the solid angle behind the 
two beam holes. Photon detectors were
positioned around the vacuum decay region,
the spectrometer, and the calorimeter to veto particles escaping the fiducial 
region of the detector. 

The muon system was located downstream of the calorimeter, 
shielded by 10 cm of lead followed by 4m 
of steel. Behind the steel was a plane of muon hodoscopes, consisting of 
15cm wide scintillator paddles oriented vertically. Behind this 
hodoscope was another meter of steel, followed by two more planes of 
scintillator paddles, one oriented vertically and one horizontally. 
 
  The  hardware trigger for this analysis  
required at least one hit in the last two banks of muon counters and 
 at least three energetic in-time clusters in the CsI
 calorimeter. The Level 3 software trigger required two tracks which 
formed a good vertex,  with one  
one track having an $E/p$ value greater than 0.7, 
consistent with an electron. 
More detail of the 
KTeV detector can be found in \cite{ed}.

A detailed Monte Carlo simulation was used to study
detector performance and acceptance, to simulate  backgrounds, and to select 
cuts.  For the LFV decays, a uniform phase space decay distribution was assumed. 

  The number of $K_L$ decays in our fiducial volume, which we refer to as 
the flux, was 
determined for each decay mode by comparison to a
similar decay with a well-known branching fraction. 
Using a normalization mode similar to the signal mode cancels many systematic
uncertainties. For the decay $K_L \rightarrow \pi^0 \mu^{\pm} e^{\mp}$, the 
normalization mode 
was $K_L \rightarrow \pi^+\pi^-\pi^0$. For $K_L 
\rightarrow \pi^0 \pi^0
\mu^{\pm} e^{\mp}$ and $\pi^0 \rightarrow \mu^{\pm} e^{\mp}$, the
normalization mode was $K_L \rightarrow \pi^0 \pi^0 \pi^0_D$, where $\pi^0_D$ 
denotes a $\pi^0$  Dalitz decay, $\pi^0 \rightarrow e^+e^- \gamma$. For 
all values of the flux and single event sensitivity quoted below, the systematic error 
was determined by varying the analysis cuts and noting the change in the measured flux.  
An additional 2\% systematic error on 
the efficiency of the muon trigger was included, since there was  no muon 
requirement for either normalization mode. 
The uncertainty in the branching fraction of the normalization modes
was included as a systematic error.

We first consider 
  the decay $K_L \rightarrow \pi^0 \mu^{\pm} e^{\mp}$. The signature for this 
 decay was two charged tracks (one electron 
and one muon) and two neutral clusters.
 The charged tracks were required to form a good vertex within the 
fiducial decay volume,and both tracks were required 
to match a cluster in the CsI calorimeter. One charged track was
required to have an $E/p$ ratio within 5\% of 1.0 and a 
transverse shower shape consistent with an 
electromagnetic shower. 
 A loose cut on the TRD information (98\% efficient for electrons) 
gave an additional cross-check on electron identification.
The second track was
required to deposit less than 1 GeV of energy in the calorimeter, 
consistent with a 
minimum ionizing muon, and to have a momentum greater than 8 GeV/c. 
 The projection of the downstream segment of the muon track was also 
required to match hits in all three  hodoscope planes of the muon 
detector, within a road 
determined by the expected multiple scattering. 

 The $\pi^0$ was reconstructed by its decay to two photons which were
detected as clusters in the calorimeter with no associated charged tracks 
and with transverse shower shapes consistent with an electromagnetic shower. 
The energy and position of the neutral 
clusters along with the location of the charged vertex were used to
calculate $M_{\gamma \gamma}$, the invariant mass of the two photon system. 
$M_{\gamma \gamma}$ was required to be within 1.4 $\sigma$ of the $\pi^0$  
mass, where $\sigma$ is the $\pi^0 $ mass resolution of 1.4 MeV/c$^2$, as
determined from the normalization mode. This requirement was chosen to 
optimize the ratio $S/\sqrt{B}$, where $S$ is the number of signal events and $B$ is 
the number of background events.

   The following kinematic cut further reduced backgrounds. 
Assuming a signal mode decay, we calculated the square of the $\pi^0$ momentum in 
the $K_L$ rest frame. For many backgrounds this quantity has an 
unphysical negative value.  We required this quantity to lie between 0 and 0.025 
(GeV/c)$^2$, where the upper value is the kinematic cutoff in the signal mode.  

\begin {figure}
\includegraphics  [width=3.3in] {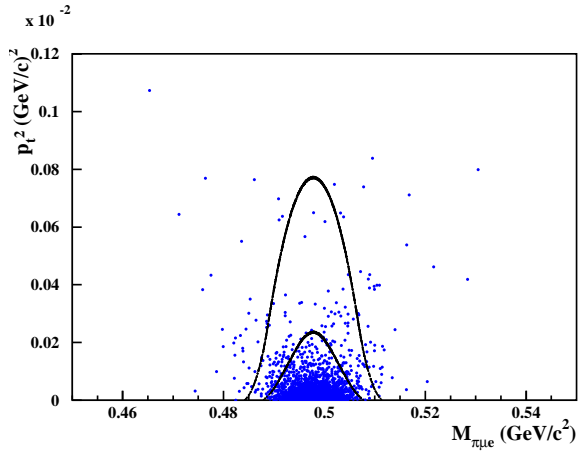}
\caption{Signal Monte Carlo events for the decay $K_L \rightarrow \pi^0 \mu^{\pm} 
e^{\mp}$  in the 
$p_t^2-M_{\pi^0 \mu e}$ plane. All cuts except the signal region cut have been made. 
The  inner contour shows the signal region, and the outer contour indicates the 
control region. } 
\label{signal}
\end{figure}

   The  flight direction of the parent $K_L$ can be approximated by a line 
from the center of 
the target to the decay vertex. We defined $p_t$ to be the sum of the 
momentum components of all
final-state particles perpendicular to this direction.  For well-reconstructed 
signal events $p^2_t$ should be
close to zero.  The signal and control regions were defined using a likelihood variable 
{\bf {\em L}} derived from  $p_t^2$ and   $M_{\pi^0 \mu e}$, the invariant mass of 
the $\pi^0 \mu e$ system, in the following way.      
  Using signal Monte Carlo, the 
$K_L$ mass distribution was fit with a Gaussian, and
the $p_t^2$  distribution was fit with a three-component exponential, producing 
probability density functions (PDFs) for these variables. Since these variables were 
found to be uncorrelated, the joint PDF was defined  
as the product of the two single-variable PDFs. Then {\bf {\em L}} was calculated for 
each event by evaluating the joint PDF at the $p_t^2$ 
and $M_{\pi^0 \mu e}$ value for that event. 
The signal (control) region was defined by a cut on {\bf {\em L}} chosen to retain 
95\%  (99\%) of signal 
Monte Carlo events after all other cuts were applied. Both the signal and 
control regions were blind during the analysis.  
  Figure \ref{signal} shows the $p_t^2-M_{\pi^0 \mu e}$  plane with 
$K_L \rightarrow \pi^0 \mu^{\pm} e^{\mp}$ signal Monte 
Carlo events shown as points, and the signal and control regions shown as 
solid
contours. 

The dominant background for $K_L \rightarrow \pi^0 \mu^{\pm} e^{\mp}$  was 
the decay $K_L \rightarrow 
\pi^{\pm}e^{\mp}\nu_e$ ($K_{e3}$), with a $\pi^{\pm}$ decay or punch through
to the muon hodoscopes, 
accompanied by two accidental photons faking a 
$\pi^0$. Since accidental photons were often accompanied by other 
accidental 
activity, we made stringent anti-accidental cuts to reduce this background. 
An event was cut if any additional charged
tracks were present. We allowed no extra in-time hit pairs in the drift chambers 
upstream of 
the analysis magnet  and at most two extra in-time pairs downstream of 
the magnet. 
We also cut on the number of partial track stubs in 
the upstream chambers. No more than 300 MeV
of energy could be present in any of the photon veto counters surrounding
the vacuum decay region, the drift chambers, and the calorimeter. The energy 
deposited in the BA calorimeter was required to be less than 15 GeV to veto events 
in which an energetic
photon escaped through one of the beam holes.   

\begin {figure}
\includegraphics  [width=3.in] {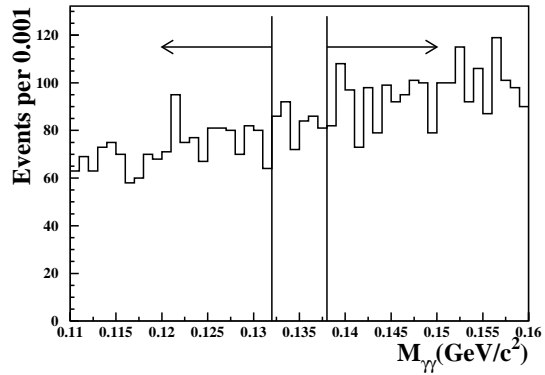}
\caption{ $M_{\gamma \gamma}$ distribution for $K_L \rightarrow \pi^0 \mu^{\pm}e^{\mp}$ 
search data, for events outside the 
signal and control regions, with all cuts 
in place except the $M_{\gamma \gamma}$ cut. The arrows show the regions 
used for the sideband background estimate. }
\label{mgg}
\end{figure}

  Figure \ref{mgg} shows the $M_{\gamma \gamma}$ distribution
for data outside the signal and control regions, with all cuts applied except the 
$M_{\gamma \gamma}$ cut.  This smooth distribution shows no peak at the $\pi^0$ mass. 
We therefore used the $M_{\gamma \gamma}$  sidebands above and below the $\pi^0$ mass 
region (0.11 GeV/$c^2< M_{\gamma \gamma}< $0.132 GeV/c$^2$ and 0.138 GeV/$c^2< M_{\gamma 
\gamma} <$ 0.16 GeV/c$^2$), but inside 
the signal or control regions in {\bf {\em L}}, to estimate the $K_{e3}$ backgrounds.  
The $K_{e3}$ background was thus estimated to be 
  0.56 $\pm 0.23$  events in the signal region and 2.56 $\pm 0.49$ events 
in the control region.

   A second source of background was $K_L \rightarrow 
 \pi^0\pi^{\pm}e^{\mp}\nu_e$ ($K_{e4}$), with a charged pion decay 
 or punch through.  
A kinematic cut to reduce this background was defined by
assuming a $K_{e4}$ decay and calculating 
 the magnitude of the unseen neutrino's momentum in the $K_L$ rest 
frame.  For $K_{e4}$ 
decays, this quantity must be  positive, while for signal decays it
 is usually negative. Requiring this variable to be negative  removed 
most $K_{e4}$ background.  The remaining  $K_{e4}$  contribution was 
determined from Monte Carlo simulation to 
be $0.10  \pm 0.050$  events in the signal region and $ 1.65  \pm 0.20$ 
events in the 
control region.  Note that the $K_{e4}$  and $K_{e3}$ backgrounds must be 
added,  
since $K_{e4}$ decays do not contribute to 
the $K_{e3}$ sideband background estimate.
 
  Another possible source of background was $K_L \rightarrow \pi^+\pi^-\pi^0$ 
decays. These decays could fake the signal if one charged pion decayed to a muon
and the second was mistaken for an electron in the calorimeter and 
TRDs. However, due to the incorrect mass assignments,  $M_{\pi^0 \mu e}$ 
reconstructed about 50 MeV/c$^2$ below the true $K_L$ mass, with no tail  
extending near the 
 signal region. The $\pi /e$ rejection from both the calorimeter and the 
TRDs suppress 
 this background to a negligible level, as confirmed by both Monte Carlo 
simulation and $K_L \rightarrow \pi^+ \pi^- \pi^0$ decays 
in data from a minimum-bias trigger.

  Other sources of background  were considered but found to be negligible.   
We find an  expected total background of 0.66 $\pm 0.23$ events in the 
signal region and 
4.21 $\pm 0.53$ events in the control region. 

The signal  acceptance for $K_L \rightarrow \pi^0 \mu^{\pm}e^{\mp}$ was 
determined from Monte Carlo simulation to 
be 3.95\% for  the 1999 data and 3.91\% for the 1997 data. The total 
number of $K_L$ decays 
in the fiducial  region was determined from the normalization mode to be 
$(6.17 \pm 0.31) \times 10^{11}$, and  
the single event sensitivity (SES) for the combined data set was $(4.12 
\pm 0.21)\times 10^{-11}$ \cite{ses}.

\begin {figure}[h]
\includegraphics  [width=3.3in] {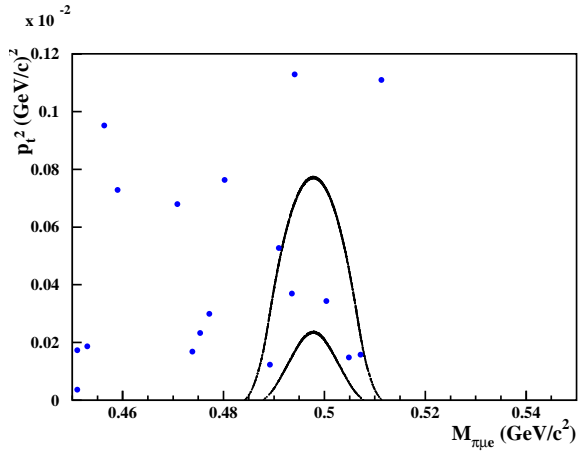}
\caption{ Surviving events in the $p_t^2-M_{\pi^0 \mu e}$  plane for the 
$K_L \rightarrow \pi^0 \mu^{\pm}e^{\mp}$ search data. The signal and 
control regions are shown as the inner and outer solid contours. }
\label{result}
\end{figure}

 When we opened the blind regions, we found 0 events in the signal region 
and 5 events in the control region, consistent with background estimations. 
Figure \ref{result} shows the $p_t^2-M_{\pi \mu e}$ plane, with the surviving 
events shown as solid dots and the signal and control region shown as contours. 

The 90\% confidence level (CL) upper limit was determined for all modes 
in the following way. 
 We stepped through a range of possible branching  fractions, using a 
Monte Carlo 
simulation to 
produce a Poisson distribution at each value.  The 
errors on the SES and backgrounds were taken into account by 
allowing these quantities to vary as Gaussian distributions with widths 
equal to their errors.  
 The resulting Poisson distributions were then used to construct 
 confidence bands, using the Feldman-Cousins prescription \cite{fc}. 
From these confidence bands we determined  $BR(K_L \rightarrow \pi^0 \mu^{\pm} 
e^{\mp}) < 7.56 \times 10^{-11}$ at the 90\% CL.  This result  represents a factor 
of 82 improvement over the previous best 
limit for this mode. \cite{old1}

  We now consider the decay $K_L \rightarrow  \pi^0 \pi^0 \mu^{\pm} 
e^{\mp}$.  The addition of a second $\pi^0$ greatly reduces the backgrounds, 
so we were able to relax some cuts to improve the signal acceptance. 
Since $K_L \rightarrow \pi^0\pi^+\pi^-$ is not a background for this mode, we did not
make a TRD requirement on the electron track, and there was no cut on the number of partial 
track stubs. We allowed up to two extra 
in-time hits in both the upstream and downstream  drift chambers.

Since we have two neutral pions in this decay, we can determine a 
neutral vertex 
independently of the charged vertex. We  required that the 
difference between the neutral and charged vertices be less than 2.5 meters. 
In addition, we  calculated an average vertex from the neutral and charged 
vertices, and 
recalculated $M_{\gamma \gamma}$ using the average vertex. The resulting 
values were 
required to lie in the region 0.132 GeV/c$^2 < M_{\gamma \gamma }<0.138$ 
GeV/c$^2$. 
Additionally, a kinematic cut on the square of the $\pi^0$ momentum in the 
$K_L$ rest frame was made on both $\pi^0$s. 

 One important source of background for this mode was the decay $K_L 
\rightarrow \pi^0 \pi^0\pi^0_D$. One electron 
could be mistaken for a muon it was mismeasured in the calorimeter and if an 
accidental muon fired the appropriate
muon hodoscope paddles. To suppress this background, we made a loose cut on the
TRD information for the muon track which rejected 85\% of 
all electrons.  This cut effectively eliminated $K_L 
\rightarrow \pi^0 \pi^0 \pi^0_D$ background. 

\begin {figure}
\includegraphics  [width=3.in] {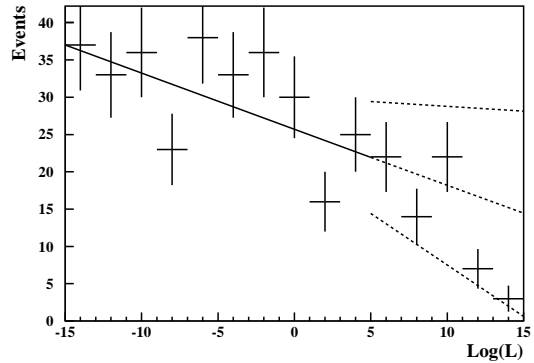}
\caption{The log({\bf {\em L}}) distribution for $K_L \rightarrow \pi^0 
\pi^0 \mu^{\pm}e^{\mp}$ search data. 
The three cuts sets as described in the text have been removed. 
A linear fit over the region -15$<$log({\bf{\em L}})$<$5 was extrapolated  
into
the signal (log({\bf{\em L})}$>$10) and control 
($5<$log({\bf{\em L}})$<10$) regions to estimate the background.   
The upper 
and lower dashed lines indicate the error bands used to assign a 
systematic error to the background estimate. }
\label{extrap}
\end{figure}

  Other backgrounds arose from $K_{e3}$ or $K_{\mu 3}$ decays with four 
accidental photons.  The $M_{\gamma \gamma}$ sidebands could not be used in this 
case to estimate the background, since they did not have a smooth distribution. The 
 background estimate was obtained instead by the extrapolation of   
a linear fit to the log({\bf {\em L}}) distribution from 
outside the control region into the 
signal and control regions.
However, when all cuts were applied, 
there were not enough events 
remaining to make a reliable extrapolation. We therefore defined three 
independent cut sets (kinematic cuts, particle ID cuts, and anti-accidental 
cuts). When we removed all three sets, we had sufficient events to make an 
extrapolation into the signal region, as shown in figure \ref{extrap}.   
After the extrapolation, we apply the suppression factor 
associated with each cut set, as determined from the data. We verified from the 
data (by applying the cut sets in various  combinations) that the three 
sets were 
indeed independent, so that we could multiply the three separate suppression 
factors to get the final background estimate. 
The total number of background events was thus estimated to be $0.44 \pm 
0.23$ in the signal 
region and  $0.43 \pm 0.17$ in the control region.  Due to the 
uncertainties in
both the extrapolation in log({\bf {\em L}}) and the suppression factors, we 
assign a systematic error on the background estimate by allowing the fit 
parameters to 
vary by 2.5 $\sigma$ from their central values. 

  The signal acceptance was 2.04\% for the 1999  data and 1.95\% for the 
1997 data. The total number of $K_L$ decays was $(6.36 \pm 0.24) 
\times 10^{11}$.  The SES for the 
combined data set was $(7.88 \pm 0.28) \times 10^{-11}$.  
 When  the blind regions were opened, we found no events in either the signal or control 
regions. We set the  90\% CL limit 
$BR(K_L \rightarrow \pi^0 \pi^0 \mu^{\pm} e^{\mp}) < 1.64 \times 10^{-10}$, which is the 
first 
limit reported for this decay.

 The search for $\pi^0 \rightarrow \mu^{\pm} e^{\mp}$,  tagged from $K_L 
\rightarrow \pi^0 \pi^0 
\pi^0$ is identical to  the $K_L \rightarrow \pi^0 \pi^0 \mu^{\pm} 
e^{\mp}$  search with the additional requirement
that $M_{\mu e}$  be in the $\pi^0$ mass 
region.   The background 
was estimated from both  $K_L \rightarrow \pi^0 \pi^0 \pi^0_D$  Monte 
Carlo and from an extrapolation of the 
log({\bf {\em L}})
 distribution  into the signal region as was done for 
$K_L \rightarrow \pi^0 \pi^0 \mu^{\pm} e^{\mp}$. 
The two methods gave  consistent results, yielding a background estimate of 
0.03 $\pm  0.015$ events in the signal region and an identical value in the 
control region. The flux for this mode was determined from ($K_L$ decays) $\times 3 
\times BR(K_L \to \pi^0 \pi^0 \pi^0)$, yielding a SES of $(1.48 \pm 0.059)\times 
10^{-10}$.  
  When  the blind regions were opened, we found no 
events in either the signal or control regions. We  
set the  90\% CL limit 
$BR(\pi^0 \rightarrow \mu^{\pm} e^{\mp}) < 3.59 \times 10^{-10}$.
Our limit on $\pi^0 \rightarrow \mu^{\pm} 
e^{\mp}$ is equally sensitive to both charge modes, while the previous best limits were not 
\cite {zeller1},\cite{zeller2}. Assuming equal contributions from both charge combinations, 
our result is about a factor of two better than the previous best limit on $\pi^0 \to 
\mu^+e^-$ and about a factor of 10 greater than the previous best limit on $\pi^0 \to 
\mu^-e^+$.

Although no evidence for these flavor-violating modes has been found, the pursuit 
should not dropped.  Given that we find negligible backgrounds, our techniques
could clearly be extended to  higher intensity neutral kaon beams. 

We gratefully acknowledge the support and effort of the Fermilab
staff and the technical staffs of the participating institutions for
their vital contributions.  This work was supported in part by the U.S.
Department of Energy, The National Science Foundation, The Ministry of
Education and Science of Japan,
Funda�o de Amparo a Pesquisa do Estado de S� Paulo-FAPESP,
Conselho Nacional de Desenvolvimento Cientifico e Tecnologico-CNPq and
CAPES-Ministerio Educao.


\begin {thebibliography} {99}
\bibitem {landsberg}L. G. Landsberg,  Phys. Atom. Nuc. {\bf 68}, 1190 (2005). 
\bibitem {ellis} A. Belyaev et al., Eur. Phys. J. {\bf C22}, 715 (2002).
\bibitem {cahn} R. N. Cahn and H. Harari, Nuc. Phys. {\bf B176}, 135 (1980). 
\bibitem {technicolor} S. Dimopoulos and J. Ellis, Nucl. Phys. {\bf B182},  
505 (1981);
T. Appelquist, N. Christensen,  M. Piai, and R. Shrock, Phys. Rev {\bf 
D70}, 093010 (2004). 
 \bibitem {ed}  A. Abouzaid et al., Phys. Rev. Lett. {\bf 99}, 081803 (2007);
A. Alavi-Harati et al., Phys. Rev. {\bf D67}, 012005 (2003);
G. E. Graham, Ph. D. Thesis, University of Chicago, 1999; 
C. Bown et al., Nucl. Instrum. Meth. {\bf A369}, 248 (1996).  
\bibitem {ses} The single event sensitivity (SES) for the 1997 and 1999 
data periods were combined as 
$SES^{-1}_{tot}=SES^{-1}_{99}+SES^{-1}_{97}.$ 
\bibitem {fc} G. J. Feldman and R. D. Cousins, Phys. Rev  {\bf D57}, 3873 
(1998). 
\bibitem {old1} K. Arisaka et al., Phys. Lett {\bf B432}, 230 (1998). 
\bibitem {zeller1} R. Appel et al., Phys. Rev. Lett. {\bf 85}, 2450 (2000).
\bibitem {zeller2} R. Appel et al., Phys. Rev. Lett. {\bf 85}, 2877 (2000).  
\end{thebibliography}
 \end{document}